\documentclass[letterpaper]{jpconf}
\def\sqr#1#2{{\vcenter{\vbox{\hrule height.#2pt
            \hbox{\vrule width.#2pt height#1pt \kern#1pt
                  \vrule width.#2pt}\hrule height.#2pt}}}}

\def\square
 {\mathop{\mathchoice{\sqr{12}{15}}{\sqr{9}{12}}
{\sqr{6.3}{9}}{\sqr{4.5}{9}}}}

\begin{document}

\title{GLSM's, gerbes, and Kuznetsov's homological projective duality}

\author{Eric Sharpe}

\address{Physics Department, Robeson Hall (0435),
Virginia Tech, Blacksburg, VA  24061}

\ead{ersharpe@vt.edu}

\begin{abstract}
In this short note we give an overview of recent work on string propagation
on stacks and applications to gauged linear sigma models.  We begin by
outlining noneffective orbifolds (orbifolds in which a subgroup acts
trivially) and related phenomena in two-dimensional gauge theories, which
realize string propagation on gerbes.  We then discuss the `decomposition
conjecture,' equating conformal field theories of strings on gerbes and
strings on disjoint unions of spaces.  Finally, we apply these ideas to
gauged linear sigma models for complete intersections of quadrics, and use
the decomposition conjecture to show that the Landau-Ginzburg points of
those models have a geometric interpretation in terms of a (sometimes
noncommutative resolution of) a branched double cover, realized via
nonperturbative effects rather than as the vanishing locus of a superpotential.
These examples violate old unproven lore on GLSM's (namely, that geometric
phases must be related by birational transformations), and we conclude by
observing that in these examples (and conjecturing more generally in GLSM's),
the phases are instead related by Kuznetsov's `homological projective duality.'
(Submitted to the proceedings of the conference {\it Quantum Theory and 
Symmetries 6} (Lexington, Kentucky, July 2009).)
\end{abstract}

\section{Introduction}

In this short note we will outline a few of the results in
\cite{nr,msx,glsm,hhpsa,cdhps}.  In general terms, these papers
outline the definition and some applications of string compactifications
on (smooth, Deligne-Mumford) stacks and, in particular, gerbes.

The original motivation for understanding string compactifications on
stacks was to investigate what new string vacua, new conformal field
theories, arise from the near-geometries provided by stacks, and to
better understand some existing theories that were already special
cases of string compactifications on stacks.

This note will focus on the special case of stacks that are gerbes.
A string on a gerbe is described by a noneffective orbifold (an
orbifold in which a subgroup acts trivially on the space) or
noneffective gauge theory.
Despite the fact that the group action is trivial, physics nevertheless
knows about that (trivially-acting) subgroup
via nonperturbative effects, as we shall outline in section~\ref{noneff-orb}.

Massless spectra resulting from such noneffective group actions appear
to violate cluster decomposition, a foundational property of most quantum
field theories, but these theories are nevertheless nontrivial
because they are equivalent to sigma models on disjoint unions of ordinary
spaces, a result known as the `decomposition conjecture,' and which we
review in section~\ref{decomp}.

The decomposition conjecture has several applications,
including 
examples in Gromov-Witten theory and the geometric Langlands program.
In section~\ref{app-glsms}, 
we focus on a different application of the decomposition conjecture,
namely
to understand the Landau-Ginzburg points of
certain puzzling gauged linear sigma models.  We point out that the
Landau-Ginzburg points in question have, for the most part,
a ${\bf Z}_2$ gerbe structure to which the decomposition conjecture
can be applied,
and find a
mathematically sensible description of the Landau-Ginzburg points in
terms of branched double covers (the double cover arising from the
application of the decomposition conjecture to a ${\bf Z}_2$ gerbe).
The resulting theories realize geometry from GLSM's in a novel fashion,
and also violate the lore that all geometric phases in GLSM's are birational
to one another.  In examples, we observe that geometric phases are related
by Kuznetsov's `homological projective duality' (instead of
`birationality'), and conjecture that this is true
more generally.

\section{Noneffective orbifolds and gauge theories}   \label{noneff-orb}

This note will concern applications of noneffective orbifolds to
physics and geometry, so let us beginning by defining a noneffective
orbifold.  It is a quotient $[X/G]$, where a subgroup of $G$, call it $K$,
acts trivially on $X$.  Such quotients are also examples of $K$-gerbes.

To understand why such a quotient is physically different from
the corresponding effective orbifold $[X / (G/K) ]$,
let us work through an example.
Consider the orbifold $[X/D_4]$, where the ${\bf Z}_2$ center of 
$D_4$ acts trivially on $X$.  Now, $D_4$ can be described as the
central extension
\begin{displaymath}
1 \: \longrightarrow \: {\bf Z}_2 \: \longrightarrow \: D_4 \:
\longrightarrow \: {\bf Z}_2 \times {\bf Z}_2 \: \longrightarrow \: 1
\end{displaymath}
so that 
only the ${\bf Z}_2 \times {\bf Z}_2$
acts nontrivially.
We shall see that the one-loop partition function of $[X/D_4]$ is
very different from the one-loop partition function of $[X/{\bf Z}_2 
\times {\bf Z}_2]$, thus making it clear that there is a difference.

We can enumerate the elements of $D_4$ as
\begin{displaymath}
D_4 \: = \: \{ 1, z, a, b, az, bz, ab, ba=abz \}
\end{displaymath}
where $z$ generates the ${\bf Z}_2$ center, and the elements of
$D_4/{\bf Z}_2 = {\bf Z}_2 \times {\bf Z}_2$ are
\begin{displaymath}
{\bf Z}_2 \times {\bf Z}_2 \: = \: \{ 1, \overline{a}, \overline{b},
\overline{ab} \}
\end{displaymath}
where the overline indicates the coset formed through multiplication by $z$.
Now, the one-loop partition function of $[X/D_4]$ has the form
\begin{displaymath}
Z(D_4) \: = \: \frac{1}{|D_4|} \sum_{g,h \in D_4; gh = hg}
Z_{g,h}
\end{displaymath}
Each of the $Z_{g,h}$ twisted sectors that appears, is the same
as a ${\bf Z}_2 \times {\bf Z}_2$ twisted sector (since the ${\bf Z}_2$
center of $D_4$ acts trivially), appearing with multiplicity
$|{\bf Z}_2|^2 = 4$, except for the sectors
\begin{displaymath}
{\scriptstyle \overline{a} } \square_{ \overline{b} }, \: \: \:
{\scriptstyle \overline{a} } \square_{ \overline{ab} }, \: \: \:
{\scriptstyle \overline{b} } \square_{ \overline{ab} }
\end{displaymath}
which are omitted, because the lifts of the group elements
to $D_4$ do not
commute in $D_4$.

Thus, we see that
\begin{eqnarray*}
Z(D_4) & = & \frac{
| {\bf Z}_2 \times {\bf Z}_2 | }{
| D_4 | }
| {\bf Z}_2 |^2 \left( Z( {\bf Z}_2 \times {\bf Z}_2 ) \: - \:
\mbox{some twisted sectors} \right) \\
& = & 2 \left( Z( {\bf Z}_2 \times {\bf Z}_2 ) \: - \:
\mbox{some twisted sectors} \right)
\end{eqnarray*}
In particular, the $D_4$ orbifold is significantly different from the
${\bf Z}_2 \times {\bf Z}_2$ orbifold, so we see here explicitly that
string theory does indeed know about even trivially-acting groups.

With a little more work, we can find a cleaner interpretation of this
result.
Discrete torsion acts as a sign on precisely the omitted twisted sectors,
so we see that
\begin{displaymath}
Z( [X/D_4] ) \: = \:
Z\left( [X/{\bf Z}_2 \times {\bf Z}_2] \coprod [X/{\bf Z}_2 \times {\bf Z}_2]
\right)
\end{displaymath}
with discrete torsion in one component.

Although we have focused so far on orbifolds, the same issue exists in
two-dimensional gauge theories, where it manifests as a question of whether
{\it e.g.} an abelian gauge theory with matter of charge $2$ is the
same as if matter is charge $1$.
Briefly, the perturbative physics is identical, but nonperturbatively
these two theories can differ.

For example, consider the supersymmetric ${\bf P}^{N-1}$ model, in which
the chiral superfields are taken to have charge $k$ rather than charge $1$.
In the ordinary ${\bf P}^{N-1}$ model, the axial $U(1)_A$ is
broken to ${\bf Z}_{2N}$ by instantons, whereas here it is broken to
${\bf Z}_{2kn}$.  Similarly, A model correlation functions are different.
In the ordinary ${\bf P}^{N-1}$ model, it is straightforward to show that
\begin{displaymath}
\langle X^{N(d+1)-1 } \rangle_d \: = \: q^d
\end{displaymath}
where $X$ corresponds to the generator of the classical cohomology ring.
Here, by contrast,
\begin{displaymath}
\langle X^{N(kd+1)-1 } \rangle_d \: = \: q^d
\end{displaymath}
As a result, quantum cohomology rings differ.  In the ordinary
${\bf P}^{N-1}$ model,
the quantum cohomology ring is
\begin{displaymath}
{\bf C}[x] / (x^N - q)
\end{displaymath}
whereas in the present case it is
\begin{displaymath}
{\bf C}[x] / (x^{kN} - q )
\end{displaymath}
In each case, we see meaningfully different physics.

On a compact worldsheet, the distinction above follows from the fact
\cite{ronen-jacques-priv}
that to uniquely specify a Higgs field, one must specify to which bundle
it couples.  For example, if a $U(1)$ gauge field couples to a
line bundle $L$, then to say a scalar $\phi$ has charge $Q$ means
$\phi \in \Gamma(L^{\otimes Q} )$.  Different bundles implies different
zero modes, which implies different anomalies, and hence different physics.
For noncompact worldsheets, there is an analogous argument using periodicity
of the theta angle \cite{ronen-jacques-priv}.

This phenomenon is not specific to two dimensions, but also has analogues
in four dimensions.  For example, the same effect is at the heart of
the (nonperturbative) distinction between $SU(n)$ and $SU(n)/{\bf Z}_n$
gauge theories, and the distinction between Spin$(n)$ and $SO(n)$ gauge
theories.  For ${\cal N}=1$ supersymmetry in four dimensions, 
there are results \cite{pouliot,strassler} describing Seiberg duality
between Spin$(n)$ gauge theory with massive spinors and $SO(n)$
gauge theories with ${\bf Z}_2$ monopoles.
In ${\cal N}=4$ supersymmetry in four dimensions, this phenomenon is crucial
for the physical understanding of the geometric Langlands program
\cite{kap-ed}.

\section{Decomposition conjecture}   \label{decomp}

Consider a quotient $[X/H]$, where
\begin{displaymath}
1 \: \longrightarrow \: G \: \longrightarrow \: H \: \longrightarrow \:
K \: \longrightarrow \: 1
\end{displaymath}
and $G$ acts trivially, and is a finite group.
We claim that
\begin{displaymath}
{\rm CFT}\left( [X/H] \right) \: = \:
{\rm CFT}\left( \left[ \left( X \times \hat{G} \right)/K \right] \right)
\end{displaymath}
(together with some $B$ field, as specified in \cite{hhpsa}),
where $\hat{G}$ is the set of irreducible representations of $G$.
As the right-hand side is a disjoint union, this relates strings on
gerbes to strings on disjoint unions of spaces (or non-gerbe stacks).
We call this the {\it decomposition conjecture}.

When $K$ acts trivially on $\hat{G}$, the decomposition conjecture above
reduces to the statement that
\begin{displaymath}
{\rm CFT}\left( [X/H] \right) \: = \: 
{\rm CFT}\left( \coprod_{ \hat{G} } (X, B) \right)
\end{displaymath}
where the $B$ field is determined by the image of the characteristic
class of the gerbe under the map defined by an element of $\hat{G}$:
\begin{displaymath}
H^2(X, Z(G)) \: 
\stackrel{ Z(G) \rightarrow U(1) }{\longrightarrow}
\:
H^2(X, U(1))
\end{displaymath}

The decomposition conjecture satisfies a number of checks, for example:
\begin{itemize}
\item For global quotients by finite groups, one can check explicitly that
partition functions match at arbitrary worldsheet genus.
\item This implies (by virtue of D-branes) a statement about K-theory, namely
that
\begin{displaymath}
K_H(X) \: = \: {\rm twisted}\: K_K(X \times \hat{G})
\end{displaymath}
which can be checked independently.
\item It is consistent with standard results on sheaves on gerbes,
namely that sheaves on gerbes decompose (in the same way), and that
Ext group elements between sheaves on different components vanish.
\item It implies results in Gromov-Witten theory, which are being checked
in papers including \cite{ajt1,ajt2,ajt3,t1,gt1,xt1}.
\item One can compute the Toda mirrors to Fano toric stacks, and and compare
quantum cohomology computations on either side of the duality.
\end{itemize}

The example of the previous section, $[X/D_4]$, fits into this framework:
the decomposition conjecture predicts that the CFT of $[X/D_4]$ should
look like two copies of $[X/{\bf Z}_2 \times {\bf Z}_2]$, one copy with
a flat $B$ field corresponding to nonzero discrete torsion, which is
what we found explicitly.

The decomposition conjecture also appears implicitly in the work
of \cite{kap-ed} on the physical realization of geometric Langlands;
there, after dimensionally-reducing along a Riemann surface to two
dimensions, one often finds a sigma model on a disjoint union of
moduli spaces.  This can be understood from the fact that the moduli
space of the four-dimensional theory in such cases is a gerbe, as
described in \cite{hhpsa}.

\section{Application to GLSM's}   \label{app-glsms}

Let us now apply these ideas to study gauged linear sigma models (GLSM's)
in two dimensions \cite{ed-glsm}.
In particular, let us consider the GLSM describing the complete intersection
of four quadric (degree two) hypersurfaces in ${\bf P}^7$.
This GLSM has 
\begin{itemize}
\item Eight chiral superfields $\phi_i$, each of charge 1,
corresponding to homogeneous coordinates on ${\bf P}^7$
\item Four chiral superfields $p_a$, each of charge -2,
one for each hypersurface in the complete intersection
\end{itemize}
and a superpotential
\begin{displaymath}
W \: = \: \sum_a p_a G_a(\phi) 
\end{displaymath}
where the $G_a$'s are degree two homogeneous polynomials.

Let us analyze the space of supersymmetric vacua in this theory, in
semiclassical regimes.  The D-term constraint is
\begin{displaymath}
\sum_i | \phi_i |^2 \: - \: 2 \sum_a |p_a|^2 \: - \: r \: = \: 0
\end{displaymath}
When $r \gg 0$, the $\phi_i$ cannot all vanish (from the D-terms above),
hence vanishing of F terms requires that $p_a = G_a = 0$,
and the theory appears to flow in the IR to a nonlinear sigma model
on the complete intersection ${\bf P}^7[2,2,2,2]$.

The other limit of $r$ is more interesting.
It is helpful to rewrite the superpotential as
\begin{displaymath}
W \: = \: \sum_a p_a G_a(\phi) 
\: = \: \sum_{ij} \phi_i A^{ij}(p) \phi_j
\end{displaymath}
where $A^{ij}$ is a symmetric matrix with entries linear in $p$'s,
that encodes the $G_a$'s.  Written in this form, it is clear that the
superpotential is really a mass matrix for the $\phi_i$.
In the limit that $r \ll 0$, the D terms specify that the $p_a$ cannot
all be zero, and the superpotential above appears to imply that the $\phi_i$
are all massive, suggesting that this limit flows to a nonlinear
sigma model on ${\bf P}^3$.  However, since the $r \gg 0$ limit is Calabi-Yau,
this limit should also be Calabi-Yau, a contradiction.

The correct analysis of this limit relies on two subtleties.
The first subtlety is that the $\phi_i$ are not massive everywhere,
some of their masses vanish along the locus $\{ \det A = 0 \} \subset
{\bf P}^3$.
The second subtlety is the fact that the $p$'s have nonminimal charge,
so over the part of ${\bf P}^3$ where the $\phi_i$ are all massive,
we have a nonminimally-charged abelian gauge theory, which (as outlined
previously) describes a local noneffective ${\bf Z}_2$ orbifold,
and hence a ${\bf Z}_2$ gerbe.

As a result of these subtleties and the decomposition conjecture,
physics sees the ${\bf Z}_2$ gerbe as
a double cover of ${\bf P}^3$ away from the locus
$\{ \det A = 0 \}$.
Moreover, it can be shown that there is a Berry phase around the locus
$\{ \det A = 0 \}$, which has the effect of interchangeing the two
sheets of the cover.  Thus, physics seems to be seeing a branched double
cover of ${\bf P}^3$, namely Clemens' octic double solid \cite{clemens}.  
As a consistency check, the locus $\{ \det A = 0 \}$
has the right degree for this branched double cover to be Calabi-Yau,
as one would expect of a GLSM describing a Calabi-Yau at another limit
in K\"ahler moduli space.

The result we have presented so far is noteworthy for at least two reasons:
\begin{itemize}
\item This is a novel realization of geometry in a GLSM, considering
that GLSM's ordinarily
build Calabi-Yau's as vanishing loci of potentials.  
\item This branched double cover is {\it not} birational to the 
geometry appearing at the other limit in K\"ahler moduli space,
namely ${\bf P}^7[2,2,2,2]$.  This result contradicts standard lore
in the GLSM community, namely that geometries appearing at limit points of
the same GLSM are all birational to one another.
\end{itemize}
Analogous results for nonabelian GLSM's have also been obtained in
\cite{hori-tong}.

This example, the GLSM for ${\bf P}^7[2,2,2,2]$, is just one of a number
of examples which can be analyzed in the same form.  In a subset of those
cases, including ${\bf P}^7[2,2,2,2]$, there is additional structure
to uncover, deriving from the fact that the branched double 
cover is singular, but the GLSM physics behaves as if it is smooth at those
singularities.  In such cases, when the branched double cover is
singular (a subset of all examples of the form above),
we believe the Landau-Ginzburg model is actually describing
a `noncommutative resolution' of the branched double cover worked out by
Kuznetsov.

This particular notion of `noncommutative space' is one described
by {\it e.g.}
\cite{kuz1,kuz2,kuz3,kont98,soi03,cos04,vdb1,vdb2,vdb3},
and is distinct from other notions of noncommutative space appearing
previously in the physics literature such as \cite{sw,rw}.
Specifically, a `noncommutative space' in this sense is defined via
its sheaves.

In the present case, the noncommutative resolution in question is defined
by sheaves of $B$-modules over ${\bf P}^3$, where $B$ is the sheaf of
even parts of Clifford algebras associated with the universal quadric over
${\bf P}^3$ defined by the GLSM superpotential.

Physically, those sheaves of $B$-modules are precisely \cite{kuz-priv}
the same as matrix factorizations at the Landau-Ginzburg point.
Intuitively, we can understand this result as follows.
First, let us recall matrix factorizations for a quadratic superpotential:
even though the bulk theory is massive, one still has $D0$-branes with a
Clifford algebra structure \cite{kap-li}.
In the present case, we have a Landau-Ginzburg model (obtained via a finite
amount of renormalization group flow)
fibered over ${\bf P}^3$, so a Born-Oppenheimer
analysis gives sheaves of Clifford algebras (determined by the
superpotential) and modules thereof.
This is ultimately why the D-branes precisely duplicate the definition of the
noncommutative resolution.

Thus, in addition to a novel realization of geometry, and an example in
which the two geometric limits of a GLSM are not birational to one another,
we see in addition that the GLSM is physically realizing a 
noncommutative resolution, in the sense discussed earlier.

As an aside, one way to study this noncommutative resolution is via
D-brane probes.  It can be shown \cite{addington1,addington2}
that the moduli space of
D-branes propagating on this noncommutative
resolution is a necessarily non-K\"ahler small
resolution of the singular space.  The non-K\"ahler structure makes it
essentially impossible for the D-brane moduli space to be the target of the
closed string theory, as it would break worldsheet supersymmetry.
It is allowed here because it is the open string moduli space, not where
the closed strings propagate.
(Another example where the closed string target space is different from the
D-brane moduli space is orbifolds \cite{msx}[section 8.2].
There, closed strings see an
orbifold -- a quotient stack -- whereas D-branes see a resolution
\cite{dgm}.)

Although the phases of the GLSM above are not birational,
they nevertheless do have a precise mathematical relationship:
they are related by
Kuznetsov's ``homological projective
duality'' \cite{kuz1,kuz2,kuz3}.
We 
conjecture that phases in all GLSM's are always
related by homological projective
duality.

\section{Summary}

In this note we have outlined some recent results and applications of
string compactifications on near-geometries provided by gerbes.
\begin{itemize}
\item We began by describing how physics sees
noneffective group actions, which is the physics at the heart of
string propagation on gerbes.
\item We outlined the decomposition conjecture for strings propagating on 
gerbes, which states that the CFT of a nonlinear sigma model
on a gerbe matches the
CFT of a nonlinear sigma model on a disjoint union of spaces or
non-gerbe stacks.
\item We described the
application of the decomposition conjecture to GLSM's, resulting
in novel realizations of geometry, some violations of old lore on
GLSM's, a physical realization of Kuznetsov's homological projective duality,
and a concrete realization of strings on noncommutative resolutions.
\end{itemize}

\section{Acknowledgements}

We would like to thank our collaborators on the papers reviewed here,
namely M. Ando, A. Caldararu, J. Distler, S. Hellerman, A. Henriques, 
and T. Pantev, for many useful discussions.
E.S. was partially supported by NSF grants PHY-0755614, DMS-0705381.


\begin{thebibliography}{199}

\bibitem{nr} T. Pantev, E. Sharpe, ``Notes on gauging noneffective
group actions,'' {\tt hep-th/0502027}.

\bibitem{msx} T. Pantev, E. Sharpe, ``String compactifications on
Calabi-Yau stacks,'' Nucl. Phys. {\bf B733} (2006) 233-296,
{\tt hep-th/0502044}.

\bibitem{glsm} T. Pantev, E. Sharpe, ``GLSM's for gerbes (and other
toric stacks),'' Adv. Theor. Math. Phys. {\bf 10} (2006) 77-121,
{\tt hep-th/0502053}.

\bibitem{hhpsa} S. Hellerman, A. Henriques, T. Pantev, E. Sharpe,
M. Ando, ``Cluster decomposition, T-duality, and gerby CFTs,''
Adv. Theor. Math. Phys. {\bf 11} (2007) 751-818,
{\tt hep-th/0606034}.


\bibitem{cdhps} A. Caldararu, J. Distler, S. Hellerman, T. Pantev,
E. Sharpe, ``Non-birational twisted derived equivalences in abelian GLSMs,''
{\tt arXiv:  0709.3855}.




\bibitem{ronen-jacques-priv} J. Distler, R. Plesser, private communication.

\bibitem{pouliot} P. Pouliot, ``Chiral duals of non-chiral susy gauge
theories,'' Phys. Lett. {\bf B359} (1995) 108-113,
{\tt hep-th/9507018}.

\bibitem{strassler} M. Strassler, ``Duality, phases, spinors and monopoles
in $SO(n)$ and Spin$(n)$ gauge theories,'' JHEP {\bf 9809} (1998) 017,
{\tt hep-th/9709081}.

\bibitem{kap-ed} A. Kapustin, E. Witten, ``Electric-magnetic duality and
the geometric Langlands program,'' Comm. Number Theory Phys.
{\bf 1} (2007) 1-236, {\tt hep-th/0604151}.

\bibitem{ajt1} E. Andreini, Y. Jiang, H.-H. Tseng, ``On Gromov-Witten theory
of root gerbes,'' {\tt arXiv:  0812.4477}.

\bibitem{ajt2} E. Andreini, Y. Jiang, H.-H. Tseng, ``Gromov-Witten theory
of product stacks,'' {\tt arXiv:  0905.2258}.

\bibitem{ajt3} E. Andreini, Y. Jiang, H.-H. Tseng, ``Gromov-Witten theory
of etale gerbes, i:  root gerbes,'' {\tt arXiv:  0907.2087}.

\bibitem{t1} H.-H. Tseng, ``On degree zero elliptic orbifold
Gromov-Witten invariants,'' {\tt arXiv:  0912.3580}.

\bibitem{gt1} A. Gholampour, H.-H. Tseng, ``On Donaldson-Thomas invariants
of threefold stacks and gerbes,'' {\tt arXiv:  1001.0435}.

\bibitem{xt1} X. Tang, H.-H. Tseng, ``Duality theorems of \'etale gerbes
on orbifolds,'' {\tt arXiv:  1004.1376}.

\bibitem{ed-glsm} E. Witten, ``Phases of $N=2$ theories in two dimensions,''
Nucl. Phys. {\bf B403} (1993) 159-222,
{\tt hep-th/9301042}.



\bibitem{clemens} H. Clemens, ``Double solids,'' Adv. Math. {\bf 47} (1983)
107-230.

\bibitem{hori-tong} K. Hori, D. Tong, ``Aspects of non-Abelian gauge
dynamics in two-dimensional $N=(2,2)$ theories,''
JHEP {\bf 0705} (2007) 079,
{\tt hep-th/0609032}.


\bibitem{kuz1} A. Kuznetsov, ``Homological projective duality,''
Publ. Math. Inst. Hautes \'Etudes Sci. {\bf 105} (2007) 157-220,
{\tt math.AG/0507292}.

\bibitem{kuz2} A. Kuznetsov, ``Derived categories of quadric fibrations
and intersections of quadrics,'' Adv. Math. {\bf 218} (2008) 1340-1369,
{\tt math.AG/0510670}.

\bibitem{kuz3} A. Kuznetsov, ``Homological projective duality for
Grassmannians of lines,'' {\tt math.AG/0610957}.

\bibitem{kont98} M. Kontsevich, ``Course on non-commutative geometry,''
ENS, 1998, lecture notes at
{\tt http://www.math.uchicago.edu/\verb,~,arinkin/langlands/kontsevich.ps}.

\bibitem{soi03} Y. Soibelman, ``Lectures on deformation theory and
mirror symmetry,'' IPAM, 2003, 
{\tt http://www.math.ksu.edu/\verb,~,soibel/ipam-fina.l.ps}.

\bibitem{cos04} K. Costello, ``Topological conformal field theories and
Calabi-Yau categories,'' Adv. Math. {\bf 210} (2007) 165-214, 
{\tt math.QA/0412149}.

\bibitem{vdb1} M. van den Bergh, ``Three-dimensional flops and noncommutative
rings,'' Duke Math.J. {\bf 122} (2004) 423-455.

\bibitem{vdb2} M. van den Bergh, ``Non-commutative crepant resolutions,''
pp. 749-770 in {\it The legacy of Niels Henrik Abel:  the Abel
bicentennial, Oslo, 2002}, Springer, 2004.

\bibitem{vdb3} M. van den Bergh, ``Non-commutative quadrics,''
{\tt arXiv:  0807.3753}.

\bibitem{sw}  N. Seiberg, E. Witten, ``String theory and noncommutative
geometry,'' JHEP {\bf 9909} (1999) 032, {\tt hep-th/9908142}.

\bibitem{rw} D. Roggenkamp, K. Wendland, ``Limits and degenerations of
unitary conformal field theories,'' Comm. Math. Phys. {\bf 251}
(2004) 589-643, {\tt hep-th/0308143}.


\bibitem{kuz-priv} A. Kuznetsov, private communication.


\bibitem{kap-li} A. Kapustin, Y. Li, ``D-branes in Landau-Ginzburg models
and algebraic geometry,'' JHEP {\bf 0312} (2003) 005,
{\tt hep-th/0210296}.


\bibitem{addington1} N. Addington, ``The derived category of the
intersection of four quadrics,'' {\tt arXiv:  0904.1764}.

\bibitem{addington2} N. Addington, ``Spinor sheaves on singular quadrics,''
{\tt arXiv:  0904.1766}.


\bibitem{dgm} M. Douglas, B. Greene, D. Morrison, ``Orbifold resolution
by D-branes,'' Nucl. Phys. {\bf B506} (1997) 84-106,
{\tt hep-th/9704151}.


\end{thebibliography}
\end{document}